# A Survey: Credit Sentiment Score Prediction

A. N. M. Sajedul Alam[1], Junaid Bin Kibria[1], Arnob Kumar Dey[1],
Zawad Alam[1], Shifat Zaman[1], Motahar Mahtab[1], Mohammed Julfikar Ali Mahbub[1],
Annajiat Alim Rasel[1]

[1] Department of Computer Science and Engineering, School of Data and Sciences (SDS),
BRAC University, 66 Mohakhali, Dhaka 1212, Bangladesh
{a.n.m.sajedul.alam, junaid.bin.kibria, al.hasib.mahamud, arnob.kumar.dey,
muhammed.zahidul.amin, md.sabbir.hossain1}@g.bracu.ac.bd,
annajiat@bracu.ac.bd

**Abstract.** Manual approvals are still used by banks and other NGOs to approve loans. It takes time and is prone to mistakes because it is controlled by a bank employee. Several fields of machine learning mining technologies have been utilized to enhance various areas of credit rating forecast. A major goal of this research is to look at current sentiment analysis techniques that are being used to generate creditworthiness.

## 1 Introduction

Banks and other NGOs still use manual approvals to approve loans. In this way, the bank employee has to carefully monitor whether the client is eligible for the loan or how much loan should be given to him. Since it is managed by a bank employee, it is time-consuming and prone to errors. If a client does not council their loan money then the bank suffers a loss. So loan prediction methods can be used to approve a loan in a short time and at low risk.

Individual credit scoring is an essential and effective risk management technique used by banks and other financial institutions. It makes appropriate loan issue recommendations and avoids financial risks. As a result, companies, as well as banks, are striving to combat the credit score threat with novel automated solutions in order to protect their own cash and customers. In recent years, several data mining techniques as well as machine learning approaches have been used to improve various aspects of individual credit score prediction.

Sentiment classification, often referred to as information extraction, is a text processing method for determining the appropriate intensity of a message. It is a common method used by businesses to assess and classify customer feedback on an item, business, or idea. It also includes analyzing literature for attitude and secondary opinions with machine learning, data analysis as well as artificial intelligence. Sentiment analysis tools enable businesses to harvest data from unstructured and disorganized text found on the web in areas such as blogging, mail, conversation, social platforms, reviews, service requests, and communities.

Predicting individual credit scores using sentiment analysis is a fascinating technology by which we can predict the creditworthiness of individuals by analyzing their sentiment data using sentiment analysis. On the other hand, we can also classify the loan applications regarding the sentiment for taking the decision of granting them.

In reality, the current dominance of computers demanding natural language processing approaches has created a culture that emphasizes model precision above investigating the causes of their failures. Nevertheless, feature representation is required for a variety of subsequent machine intelligence and natural language uses, such as banking, healthcare, and self-driving. Rather than presenting a "novel method," this study investigates the regression model of several popular sentiment methodological approaches in the financial sector. They demonstrated that [1] methods from the same category have identical patterns of errors as well as [2] recurrent errors having six separate sorts of language properties. The findings give useful insights and practical recommendations for the advancement of emotion detection methods in business organizations.

Creditworthiness assessment is an important step for financial firms to take when it comes to business loans. Manual study of a firm's overall status via customer-appropriate research studies, but on the other side, takes time and work. In this study, the GMKL model was applied to propose a breakthrough credit risk rating approach that automates decision-making. Due diligence reports textual information for views, mining clients produce sentiment indexes, which are subsequently used as material for model development. The approach distinguishes itself by combining sentiment classification into credit risk evaluation in an innovative manner. The method's effectiveness is evaluated using a database of real-world loan approvals.

Microfinance institutions primarily assist those who are unable to access traditional banking services due to a major absence of sufficient guarantees. Because of this, the institutions are at an extremely elevated danger. Financial firms use credit scoring to determine whether or not their customers are creditworthy. The author of this work provides a knowledge model which depicts the primary factors that might influence a credit score. A literature research was carried out to establish the many factors that influence credit scores. In the articles examined, a variety of models were discussed. The key dimensions of credit rating evaluation as well as their linkages have been identified using these models. They built a credit scoring knowledge model based on their research on current credit scoring methods in microfinance organizations. This research offered a generic model that is all-encompassing in nature. This model may also be used in the beginning stages of the rating model creation process.

Sentiment Analysis (SA) has gotten a lot of attention over the decades since it looked to be a huge advancement. Sentiment analysis (SA), is often known as opinion mining.SA is often connected with customer-voice materials, such as reviews and research replies, as well as online and web-based life and social security resources for applications ranging from education to customer service to clinical treatment. Throughout this research, they lead a Sentiment Analysis of Client Surveys that has a significant impact on a business development strategy. They used Twitter data as an input variable in the form of a CSV file. They used SVM, NB, KNN, DECISION TREE, LOGISTIC REGRESSION, RANDOM FOREST. Researchers first gather data and group it into positive & negative data categories in order to begin this procedure. After the data has been grouped, they map out the various groups. After that Researchers use KNN and other algorithms to do sentiment analysis.

Credit is essential for the existence of an online store. When a transaction is complete, each party can offer each other praise, a mid-level rating, or an unfavorable evaluation. The second evaluation is more representative of the store's actual credit rating and has the potential to successfully decrease the false brushing problem.

Peer-to-peer financing is based on an e-commerce platform and credit from online transactions. Borrowers and lenders in peer-to-peer lending can trade online using an internet platform. The transaction expenses are lower, and the financing method is quick and uncomplicated. Credit institutions currently analyze consumer credit risk using data from Facebook, LinkedIn, and Twitter users. Individuals without or with limited credit records will gain more from information sources. Creditworthiness should indeed be assessed based on an individual's overall identification, internet image as well as a business network.

Comments in code repositories represent its contributors' feelings and emotions regarding the project in that repository. For projects of organizations, it's easier to determine the emotions and feedback of providers through communications, surveys, etc. For open-source projects, it is difficult as these mediums are not available. In this paper, comments in open source project repositories were analyzed to research the emotions expressed by contributions based on various facts.

Because of media platforms and their connected apps, millions of individuals may communicate and share their ideas about something like a subject, but also display their sentiments by liking or disliking information. Several scholars have indicated a desire to use massive social information to measure, evaluate, and decide things in a broad range of domains. This strategy necessitates a set of research methodologies, the most important of which is text analytics. Sentiment analysis is the method for determining people's feelings on a topic by analyzing their online postings and activity. The next step is to separate the dualism of the posts into discrete opposing emotions such as joy, sadness, and so on. Researchers intend to address the original study of linguistic analysis by offering a revolutionary adaptive approach that combines comments from online platforms along with other big data platforms to evaluate users' perspectives along with their behaviors in the direction of an actual research direction.

TripAdvisor is a well-known platform for travel & tourism. It receives around 490 million average monthly visits, has 760 million reviews and comments, and has 8,3 million rooms, airlines, restaurants, and attractions . This travel platform has already been popular in the U.s. depending on monthly visitors from April 2018. Users on TripAdvisor may post 100-character texts and give the review a score between 1 and 5, with 1 being the lowest and 5 being the highest. This research uses sentiment analysis to rank 55 attractions in Bali depending on TripAdvisor user reviews. It classifies polarity evaluations of tourist destinations using a hybrid technique.

The sole purpose of this survey is to explore existing models of sentiment analysis that were used for generating credit scores and find out possible future implementations, limitations, and extensions of those works. In this survey, we tried to get an overall idea of a total of eight research papers to bring out an overview of implemented sentiment analysis models for predicting individual credit defaults.

## 2   Survey Details

Xing et al. [3] in their paper, it has been seen that, given the same models, sentiment analysis results in the financial sector is much less accurate. MCC falls from over 74 to over 42 on average, The F1-score drops dramatically 86 - 80.60 and accuracy falls from over 84 to over 71. The error frequencies reveal that lexicon-based systems create consistent errors, for example, SenticNet produces higher mistakes both for positively and negatively sampled data. In contrast, four/ five attempting to learn algorithms (bi-LSTM, BERT, fastText, SVM) generate more false positives than false-negative mistakes. This finding might imply that they are unable to address the problem of unbalanced data; yet, S-LSTM has learned to make well-balanced mistakes. Because of its considerable expressive capacity, BERT has the top ranking across all measures of something like the dataset of Yelp.

Inside "model clusters," pairwise correlations are greater. They find the strongest correlation between SVM and fastText in the left matrix, and so build machine learning-based clustering. Similar to the deep learning system cluster (bi-LSTM, BERT, S-LSTM), the model cluster of deep learning (S-LSTM, bi-LSTM, and BERT) does have one shade of dark. L&M, also OpinionLex were likewise significantly associated (corr with a percent of 40%), however, SenticNet loosens the lexicon-based cluster. This might be owing to the belief that SenticNet uses syntax to classify sentiment.

Zhang et al. [4] in this paper showed that the GMKL model is used in this work to propose a unique credit risk evaluation technique. This proposed method is distinct in that it includes sentiment classification with credit evaluation. When contrasted to the usual credit evaluation technique, the sentiments from over 950 research outputs of the right research report are considered as an additional reference. The sentiment score and indicators of finance are employed as input features in the GMKL model to automate the judgment process. The scientific experiment is conducted using a real-life loan-granting dataset, with over 1600 applicants serving as legitimate examples. The results suggest that integrating the emotion indicators improves classification performance significantly. Furthermore, the increase in the true negative variations that the sentiment results from the analysis of the proper research report improves the capacity to avoid fraud risk. When combining sentiment indices with financial indicators, GMKL has shown to be the best appropriate categorization model.

A novel approach has been developed by Raju K. D. and Dr. Jayasingh B. B. in order to provide accurate impressions of the restaurant or hotel. [5] Using audits, systems such SVM, KNN, NB, and so on may be set up to get results including "mad," "sad(bad)," and "glad". This is the most accurate technique to determine the survey's extreme. The feature pick has been based on the detected reviews of the specific eatery. The new data set was utilized to categorize the reviews in the Testing stage.

Authors Addi KB and Souissi N present a broad and comprehensive overview of credit scoring information in their paper. [6] With all of the existing data collecting techniques and the rate at which they are improving, financial organizations may be able to obtain substantially more useful information on customers and their environments. This study proposes a generic ontology model that may be used in any financial organization to create a generic credit scoring system.

The existing frequently used assessment method is contrasted with the evaluation system described in this study [7]. We can easily see from the comparison of images that the ratings of the evaluated comments are more comprehensive. During the upgrade, specific terms will be provided, as well as the phrase linguistic tag are more comprehensively supplied. People who are scared with compliments or even their own assessment actions are prevented from engaging.

In their work, they investigated default risks to use a publicly available dataset from PPDai [8]. Unlike American databases, its dataset is imprecise and lacks a FICO [13] system. They added behavioral patterns data into the parameter groups used to create the credit scoring model to highlight the bias of information between borrowers and lenders. Due to the fact that perhaps the dataset's length is still tiny, the experimental findings show that their model has a high degree of classification accuracy.

60425 commits of total 90 Github software projects were analyzed. SentiStrength was used for lexical sentiment extraction. This gives preset values to words inside a vocabulary that also includes popular emojis. It breaks down one sentence into small sentences and assigns scores to them. Average scores of all the small sentences are used to calculate a positive/negative score of a particular sentence. From the assigned score for all commit comments, it was seen that the average score of them was neutral. Nevertheless, ratings varied according to a computer program, day/time of the week, group dispersion, program approval, and so on. [10]

A three-stage process is used to develop a sentiment analysis model. These are dictionary construction, categorization, and predictions. For data collection, Kafka, a decentralized streaming platform, is employed. Spark is used for data processing. The researchers picked at random a sample of 600 tweets from the present Tweet set of data: 50 for every class. Every candidate's tweets were meticulously examined and classified as positive, positive, moderate, very positive, somewhat unfavorable, significantly negative, forms of prejudice, or neutral. As earlier mentioned, similar information was analyzed after textual data, text classification, stems, and numerous filtering were removed. This stage was carried out using TreeTagger, a tool for tagging texts with portion and lexical information. TreeTagger now supports negation, URLs, users, Twitter mention, hashtags, and intensifiers. Four commonly used metrics are used to assess the efficacy of a classification technique: accuracy, precision, recall, and F-score. To calculate classifier performance, the average of the four criteria for each class of both people is employed. [11]

In this study, researchers examined the overall emotion of TripAdvisor reviews, one of the most famous tourist review websites. Handhika et al. [12] used a hybrid strategy that combines two distinct techniques, the lexical-based method, and the computer learning-based method. Rather than using an inconsistent grading system, the researchers used a lexical-based approach to determine the usual label of each review. Researchers used a machine learning-based technique with specific predetermined criteria to predict the mood of existing or fresh review data. They start their sentiment analysis by scraping the date, review, star rating, and URL for several attractions that an algorithm recognizes. Researchers scan TripAdvisor reviews to determine the accuracy of the Homogeneous Ensemble Classification for each site at Bali, Indonesia, to be specific 55 noticeable interest in this experimental analysis. That reliability was determined by comparing the result of the machine learning algorithms they used, specifically the Random Forest, Bagged-Decision Tree, and Stochastic Gradient Boosting (SGB) to the label provided by the Lexical-Based Approach (SenticNet). The accuracy rate for every 55 main interests for each SGB, Homogeneous Ensemble Random Forest as well as Bagged-DT algorithm is 98.1267 percent, 95.2643 percent, and 97.9829 percent.

## 3 Analysis

We produced **Table 1** to help us analyze our survey findings by giving a brief summary of each article's core topics and classifying them into sub-domains and sub-disseminations within each domain.

**Table 1.** The following is a list of the articles that were considered for this Systematic Literature Review.

| Articles | Major Domain | Sub-Domain | Key Concept |
| --- | --- | --- | --- |
| Financial Sentiment Analysis: An Investigation into Common Mistakes and Silver Bullets Introduction [3] | Financial | Banking | In this study, they experimented with the efficiency of ML-based, lexicon-based as well as deep models for commercial sentiment analysis. By displaying mistake patterns and undertaking language analysis, they went beyond merely comparing model metrics. |
| Can sentiment analysis help mimic the decision-making process of loan granting? A novel credit risk evaluation approach using GMKL model [4] | Financial | Banking | This paper approached the risk evaluation using a model named GMKL. By using sentiment analysis, they somehow mimicked the decision making system while granting the loan |
| Machine Learning for Sentiment Analysis for Twitter Restaurant Reviews [5] | Online Platform | Social Media | Throughout this research, they show how sentimental analysis was utilized to identify consumer reviews, including how those reviews were effectively identified using various classifier algorithoms. |
| An Ontology-Based Model for Credit Scoring Knowledge in Microfinance: Towards a Better Decision Making [6] | Financial | NGO | The research described in this study is based on an analysis of many models to determine characteristics of credit score in a microfinance environment, with the goal of developing an ontological model that depicts the dimensions that affect credit score including their interrelationships. The suggested methodology will aid such |

| | | | organizations in making decisions, particularly in evaluating loan applications |
|---|---|---|---|
| Research on Credit Evaluation Model of Online Store Based on SnowNLP [7] | Online Platform | Shops | The seller's honesty and the product's quality are reflected in the online shop credit rating. The buyer's desire to acquire is strongly influenced by his or her credit rating. The assessment results of this study are more accurate, thorough, and intuitive than the credit evaluation technique usually employed in online retailers. |
| Research on Credit Scoring by fusing social media information in Online Peer-to-Peer Lending [8] | Online Platform | Social Media | In recent years, the internet coequal to the coequal borrowing sector market in China has grown rapidly. By merging data from social media with a decision tree, they created a credit-rating system. The experimental results indicate the model's classification performance. The credit score, on either side, is just not as important as the system claims. |
| Sentiment Analysis of Commit Comments in GitHub: An Empirical Study [10] | Online Platform | Social Media | Through lexical analysis, emotions expressed by contributors in github commit comments were extracted. Additionally, the relationships between emotions expressed in the comments and different factors like time/day, programming language, team distribution etc were analyzed. It was seen that projects written in Java had more negative commit comments than others, while distributed teams seemed to have more positive comments. |

| A novel adaptable approach for sentiment analysis on big social data [11] | Online Platform | Social Media | Along with its dynamic and genuine personality, obtaining public opinion through the analysis of large social data has piqued the curiosity of many. They provide such sentiment technology that reads social networking sites' communications in real-time as well as extracts people's thoughts. The proposed technique comprises first developing a dynamic lexicon of word polarities based on a collection of keywords associated with a certain topic and then categorizing tweets by integrating extra factors that greatly increase the polarization degrees of communication. |
| --- | --- | --- | --- |
| Hybrid Method for Sentiment Analysis Using Homogeneous Ensemble Classifier [12] | Online Platform | Social Media | Among the most popular tourist review websites, researchers examined the emotion of a TripAdvisor review. This is the most common polarity used by travelers. Although there is frequent disagreement between their remarks and the polarity score, this score cannot be considered indicative of all comments because it does not reflect the whole of their comments. They employed the hybrid technique as an alternate solution to these issues, that integrates two unique methods, lexical-based as well as machine learning-based. |

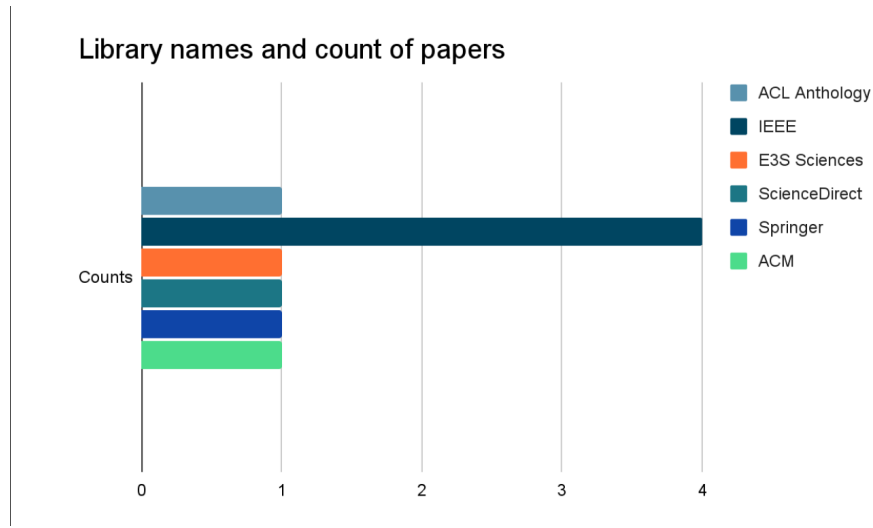

**Fig. 1.** Library names and count of papers from them.

The greatest counting number in **Fig. 1** indicates that the IEEE library holds the majority of the papers among the following papers. The number of publications in other libraries, such as ACL Anthology, E3S Sciences, ScienceDirect, Springer, ACM is the same.

Six libraries were used to uncover intriguing study topics for this project. This work selects conference and journal papers from such publications' collections that are relevant to the research issues that are being investigated in a complete literature assessment of the acceptability of sentiment analysis technology in predicticting credit scores. Using the search capabilities of each publishing repository, this study critically analyzes the publications related to our study goal. The libraries of all six publishers are included in **Table 2**.

**Table 2.** List of the libraries that were searched.

| No. | Library |
| --- | --- |
| 1 | ACL Anthology |
| 2 | IEEE |
| 3 | E3S Sciences |
| 4 | ScienceDirect |
| 5 | Springer |
| 6 | ACM |

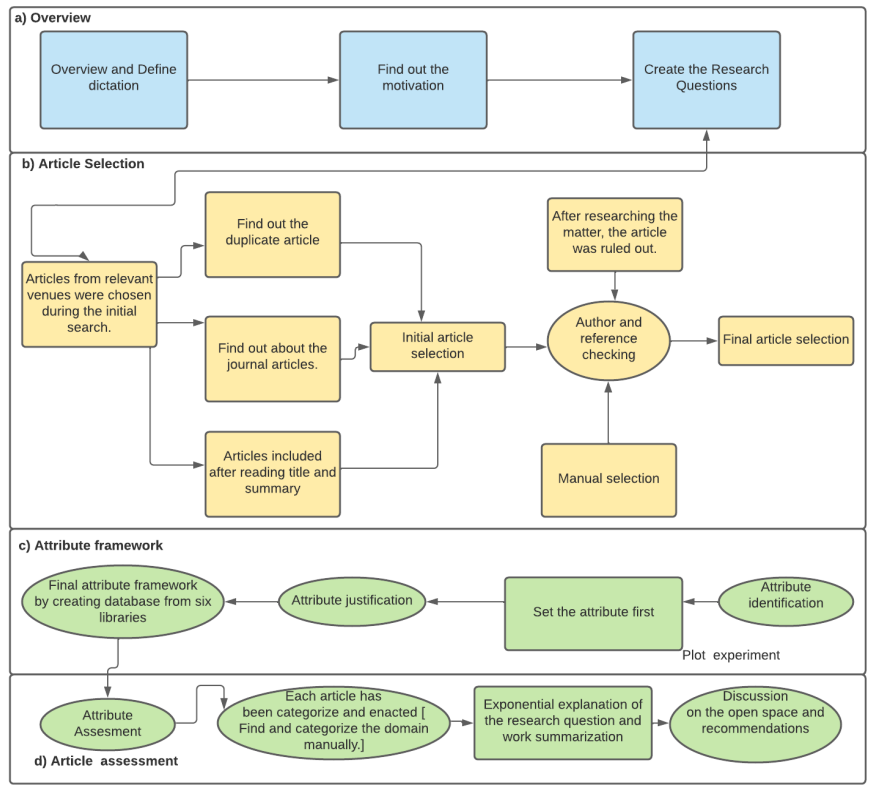

**Fig. 2.** Overview of Survey Details [2], [3], [4], [5]

The complete explanation of survey specifics is presented in **Fig. 2**, which depicts the beginning to finish a section of the survey, that is, how the articles were selected, how the survey was conducted, and how the final suggestions were discovered. Kitchenham's approach, which comprises attribute selection, attribute framework, attribute assessment, and research questions, was used to create the whole procedure shown in **Fig. 2**. [9].

### 3.1   Sub Domain List:

I.   Banking

II. Social Media
III. NGO
IV. Shops

## 3.2 Major Domain List:

I. Financial
II. Online Platform

We split the nine articles into two primary domains based on two genres: **'Financial'** and **'Online Platform'**, based on the four subdomains we picked. **Table 3** shows which papers belong to which genres, and it can be seen from the table that all of the articles in this section are allocated in 1:2 proportion.

**Table 3.** Genre distribution of selected papers.

| Genres | Titles | | Counts |
|---|---|---|---|
| **Financial** | a) | Financial Sentiment Analysis: An Investigation into Common Mistakes and Silver Bullets Introduction [3] | 3 |
| | b) | Can sentiment analysis help mimic the decision-making process of loan granting? A novel credit risk evaluation approach using GMKL model [4] | |
| | c) | An Ontology-Based Model for Credit Scoring Knowledge in Microfinance: Towards a Better Decision Making [6] | |
| **Online Platforms** | a) | Machine Learning for Sentiment Analysis for Twitter Restaurant Reviews [5] | 6 |
| | b) | Research on Credit Evaluation Model of Online Store Based on SnowNLP [7] | |
| | c) | Research on Credit Scoring by fusing social media information in Online Peer-to-Peer Lending [8] | |
| | d) | Sentiment Analysis of Commit Comments in GitHub: An Empirical Study [10] | |

| | e) | A novel adaptable approach for sentiment analysis on big social data [11] |
|---|---|---|
| | f) | Hybrid Method for Sentiment Analysis Using Homogeneous Ensemble Classifier [12] |

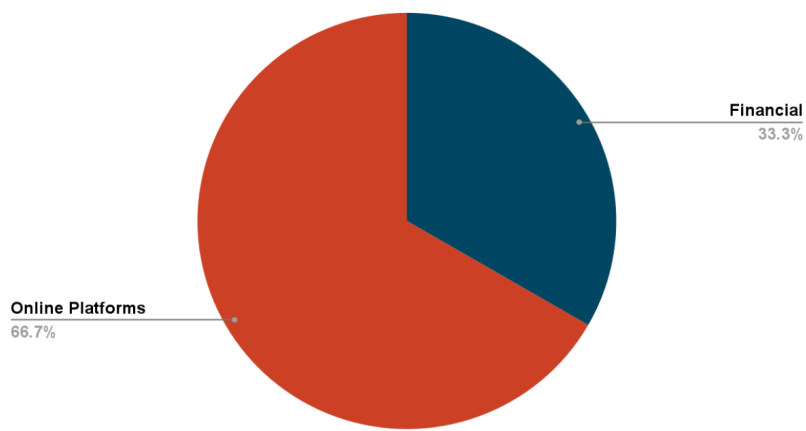

**Fig. 3.** Genre distribution of selected papers.

We have shown the categories into which we have split our works in **Fig. 3**. We discovered that one third of our documents were financial and the rest were from online platforms throughout the categorizing process, resulting in a 1:2 ratio.

## 4 Conclusion

In this paper, in total we surveyed nine articles, to be specific, three journal papers and six conference papers. We demonstrated the solutions that the surveyed papers produced to handle the problems of credit risk. We categorized the papers into two Domains including Financial and Online Platforms while further branching them into four subdomains. The findings we got from this survey are really distinguishing for the future implementations of sentiment analysis in credit scoring systems. To sum up this survey we can say that, in the banking sector, social media platforms, NGO, shops in these subdomains, there are already existing models of the sentiment credit score system and in the future more and more applications will be there to make decisions of credit worthiness of individuals by machines.